\documentclass[
  aps,
  prl,
  twocolumn,
  superscriptaddress
]{revtex4-2}

\usepackage{graphicx}
\usepackage{amsmath}
\usepackage{amssymb}
\usepackage{xeCJK}

\begin{document}
\title{Charge-Reversal-Based Extraction of Muonic Ions}
\author{Song Guo (郭松)}
\email[Corresponding author, ]{gs@impcas.ac.cn}
\affiliation{Key Laboratory of High Precision Nuclear Spectroscopy, Institute of Modern Physics, Chinese Academy of Sciences, Lanzhou 730000, China}
\affiliation{School of Nuclear Science and Technology, University of Chinese Academy of Sciences, Beijing 100049, China}
\date{August 2026}

\begin{abstract}
Negative muon capture in matter is accompanied by Auger electron emission and the formation of positively charged muonic ions. This charge reversal provides a possibility for electromagnetic extraction of the products of muon capture. Here we propose a charge-reversal-based extraction scheme for muonic ions, in which negative muons are moderated and confined in a dilute gas environment before conversion, and the resulting muonic ions are subsequently accelerated and extracted by electrostatic fields. The feasibility of the concept is investigated through order-of-magnitude estimates of the competing processes, including muon moderation, muonic-ion formation, charge exchange, and acceleration in the gas medium. A magnetic-electrostatic deceleration scheme is proposed to reduce the transverse phase space of practical low-energy muon beams while retaining controlled longitudinal motion, and is directly matched to the electrostatic capture region. The analysis indicates that muonic-ion formation and extraction within the muon lifetime are not excluded by fundamental timescale constraints, although the quantitative efficiency remains uncertain because several relevant low-energy cross sections and charge-state populations are poorly constrained experimentally.

\end{abstract}
\maketitle
Negative muons provide a unique opportunity for studying atomic and nuclear properties through the formation of muonic atoms~\cite{Knecht2020}. Owing to the much larger mass of the muon compared with the electron, muonic atoms possess significantly reduced orbital radii and enhanced sensitivities to nuclear structure and electromagnetic interactions~\cite{Anderhub1976}. These unique characteristics have enabled a broad range of applications, including precision measurements of nuclear properties, studies of strong-field atomic processes, and investigations of fundamental interactions. Recent experiments have directly demonstrated highly charged muonic atoms in low-pressure noble gases \cite{Okumura2025}. Beyond spectroscopy, the controlled production and manipulation of muonic atoms and their ionic products could open new possibilities for exploiting low-energy muon beams~\cite{Yoshida2018}.

However, converting muon capture into an extractable beam is intrinsically challenging. Efficient muon stopping favors a sufficiently dense and localized medium, whereas the same environment promotes charge exchange and neutralization of the muonic ions formed after capture. Reducing the density alleviates these collisional losses but increases the stopping length and reduces the stopping efficiency within a finite volume. Strong electric fields can shorten the extraction time, but may interfere with the controlled slowing of negative muons toward the capture regime, while slow extraction incurs additional losses from muon decay. These competing requirements are further aggravated by the large energy and spatial spreads of practical muon beams. A scheme that simultaneously accommodates muon stopping, muonic-ion formation, and rapid extraction is therefore highly desirable.

In this work, we exploit the charge-state evolution associated with muon capture rather than treating capture merely as the endpoint of the stopping process. Following capture, the muonic cascade and electronic rearrangement can remove a substantial fraction of the atomic electrons through Auger emission, producing positively charged muonic species,
\begin{equation}
\mu^- + A^0
\rightarrow
(\mu A)^{q+}+(q+1)e^- ,
\end{equation}
where $q$ is determined by the number of atomic electrons removed during the cascade~ \cite{Ninomiya2021,Vogel1980,Kirch1999,Hauser1998,Gorelkin1996,Gorelkin1997}. 

\begin{figure}[!htbp]
	\centering
	\includegraphics[width=0.5\columnwidth]{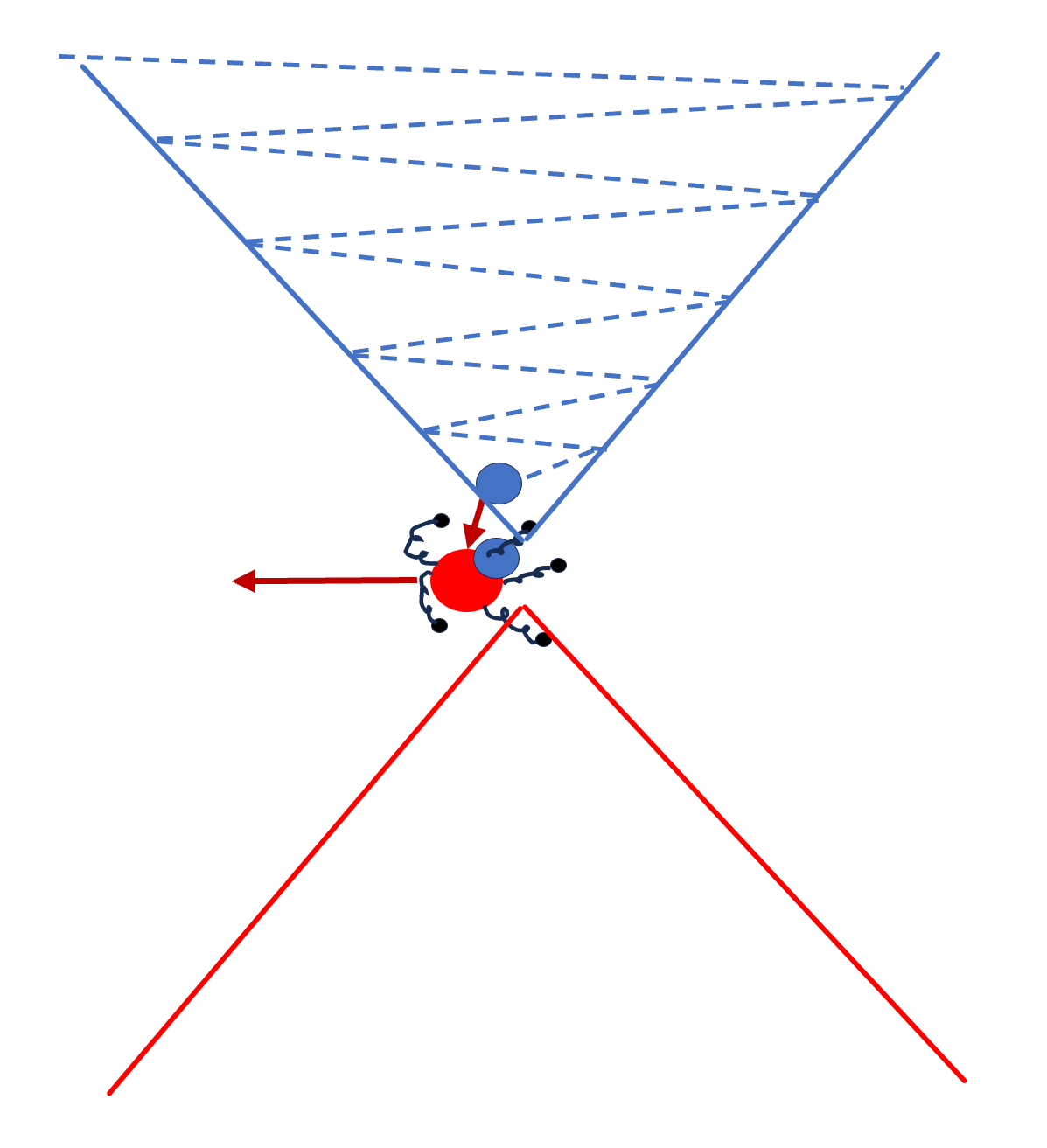}
\caption{Principle of charge-reversal-based extraction. The same electrostatic potential structure acts as a confining potential for the incident negative muon and, after charge reversal during muon capture, as an extracting potential for the resulting positively charged muonic ion.}
\label{fig:potential}
\end{figure}

This charge reversal provides a natural transition from confinement to extraction: an electrostatic potential well can confine and decelerate the incident \(\mu^-\) toward the low-energy capture regime, while the resulting positively charged muonic ion experiences the opposite force in the same potential structure and is accelerated toward the extraction region. The charge-state change induced by capture therefore automatically converts the same electromagnetic configuration from a confining potential into an extracting potential, without requiring an external switching mechanism. This principle is illustrated in Fig.~\ref{fig:potential}.

The essential requirement is that the charged muonic atoms must be extracted before their charge state is significantly modified by subsequent collisions. Therefore, the relevant condition is

\begin{equation}
t_{\rm extract}
<
t_{\rm charge\ exchange},
\end{equation}

rather than the establishment of a stable charge equilibrium. The proposed approach thus relies on preserving and utilizing the transient charge distribution generated during muonic atom formation.

The extraction field and gas density are therefore the two principal control parameters. A sufficiently strong electric field is required to rapidly accelerate and extract the positively charged muonic ions before repeated charge exchange drives them toward lower charge states. A low gas density is favored for the same reason, as it reduces the frequency of charge-exchange collisions during transport. These requirements are opposed by two additional constraints: increasing the reduced electric field enhances electron multiplication and may lead to gas avalanche or breakdown, while reducing the gas density increases the muon slowing-down time and may prevent efficient moderation within the muon lifetime. The accessible operating region is therefore determined by the intersection of these competing constraints.

To avoid excessive electric-field enhancement and surface-induced breakdown, the local electric field should remain below a characteristic upper limit of order $10^{7}\,\mathrm{V\,m^{-1}}$. At the same time, the gas density must be sufficiently high to slow the incident muons to the capture regime within their lifetime. The stopping-time estimate gives a characteristic lower bound of order $N\sim10^{21}\,\mathrm{m^{-3}}$ for moderation on a timescale of $1\,\mu\mathrm{s}$.

\begin{figure}[!htbp]
	\centering
	\includegraphics[width=\columnwidth]{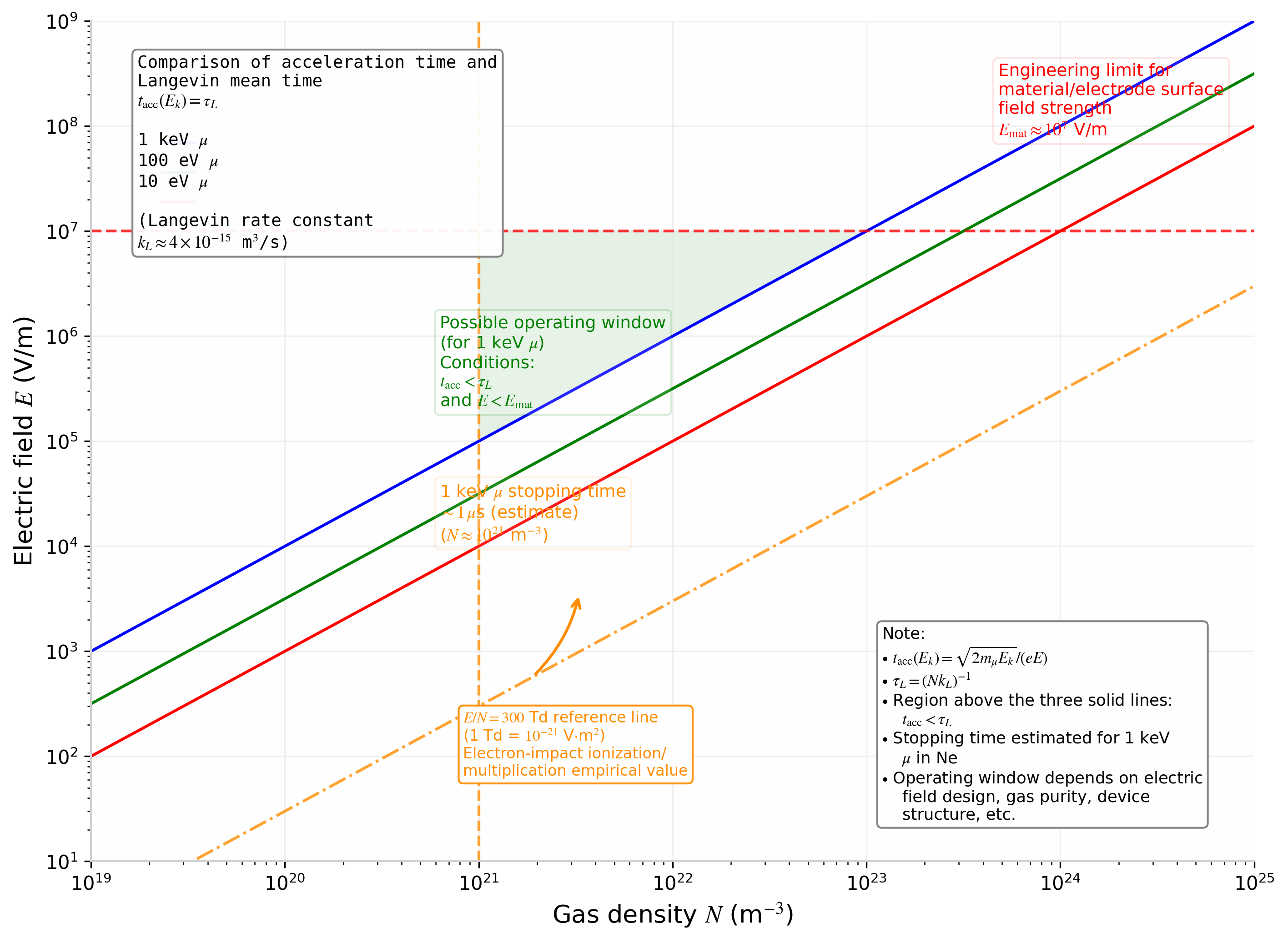}
\caption{Operating constraints in the electric-field--gas-density parameter space. The boundaries are set by the maximum electric field allowed by surface breakdown and by the requirement that muon moderation be completed within $1\,\mu\mathrm{s}$. The three diagonal boundaries correspond to the conditions for a $7+$ muonic ion in neon to reach $10$, $100$, and $1000\,\mathrm{eV}$ before a charge-exchange collision. The red line indicates $E/N=300\,\mathrm{Td}$, adopted as a conservative limit for electron multiplication.}
\label{fig:EN}
\end{figure}

A further constraint is imposed by charge exchange during extraction. As the muonic ion is accelerated, its charge-exchange cross section decreases with increasing energy. We therefore compare the acceleration time required for a muonic ion to reach a given energy with its mean charge-exchange time. Taking neon as the stopping gas and considering an initially populated $7+$ charge state, the conditions for reaching $10$, $100$, and $1000\,\mathrm{eV}$ before a charge-exchange collision define three boundaries in the $(N,E)$ parameter space, as shown in Fig.~\ref{fig:EN}. Because both the acceleration and collision rates scale with the gas density, the corresponding boundaries are primarily determined by the reduced electric field $E/N$.

Electron multiplication provides another constraint that depends primarily on $E/N$. We adopt $300\,\mathrm{Td}$ as a representative upper limit for a conservative operating region and indicate the corresponding boundary in Fig.~\ref{fig:EN}. The comparison shows that the reduced electric field required to accelerate the muonic ion before substantial charge exchange is larger than this nominal safe value. If judged solely by $E/N$, the two requirements therefore appear incompatible.

This apparent conflict does not, however, imply that the
corresponding electric-field and density combinations are unusable. The onset of a self-sustained discharge depends not only on the reduced electric field but also on the physical length of the high-field region. For a sufficiently short gap, an electron can reach the electrode after only a limited number of collisions, even when the Townsend multiplication coefficient in an extended medium is large~\cite{Loveless2019,TownsendReview2022}.

For example, consider a $50\,\mathrm{kV}$ potential drop across a
$1\,\mathrm{cm}$ high-field region in neon at $N=10^{22}\,\mathrm{m^{-3}}$.
This corresponds to
\[
E=5\times10^{6}\,\mathrm{V\,m^{-1}},
\qquad
E/N=5\times10^{5}\,\mathrm{Td}.
\]
At this density, the characteristic atomic collision length estimated
from a geometrical cross section is of order
\[
\lambda_{\rm at}\sim 0.3\,\mathrm{mm},
\]
so that an electron can gain an energy of order
\[
eE\lambda_{\rm at}\sim1.5\,\mathrm{keV}
\]
between collisions. The electron-impact ionization cross section of
neon at this energy is of order $10^{-21}\,\mathrm{m^2}$, giving the
order-of-magnitude estimate
\[
\alpha\sim N\sigma_{\rm ion}\sim10\,\mathrm{m^{-1}},
\qquad
\alpha d\sim0.1.
\]
Thus, despite the large reduced field, the finite-length multiplication
factor remains modest, and the Townsend criterion
\[
\gamma\left[\exp(\alpha d)-1\right]<1
\]
can remain satisfied for typical $\gamma\ll1$. 

The compact trapping structure therefore imposes stringent requirements on the phase-space distribution of the incident muon beam. As the kinetic energy of the $\mu^-$ particles decreases, their probability of being captured by surrounding atoms increases. It is therefore important to avoid premature thermalization before the particles reach the designated capture region. In the present design, the electric and magnetic fields are arranged in parallel, which approximately separates the radial and axial motions of the particles. This allows the deceleration to be arranged such that at least one component of the particle velocity remains relatively high during the early stage of transport, thereby reducing the probability of premature atomic capture while the particles are progressively slowed toward the capture region.

\begin{figure}[!htbp]
	\centering
	\includegraphics[width=\columnwidth]{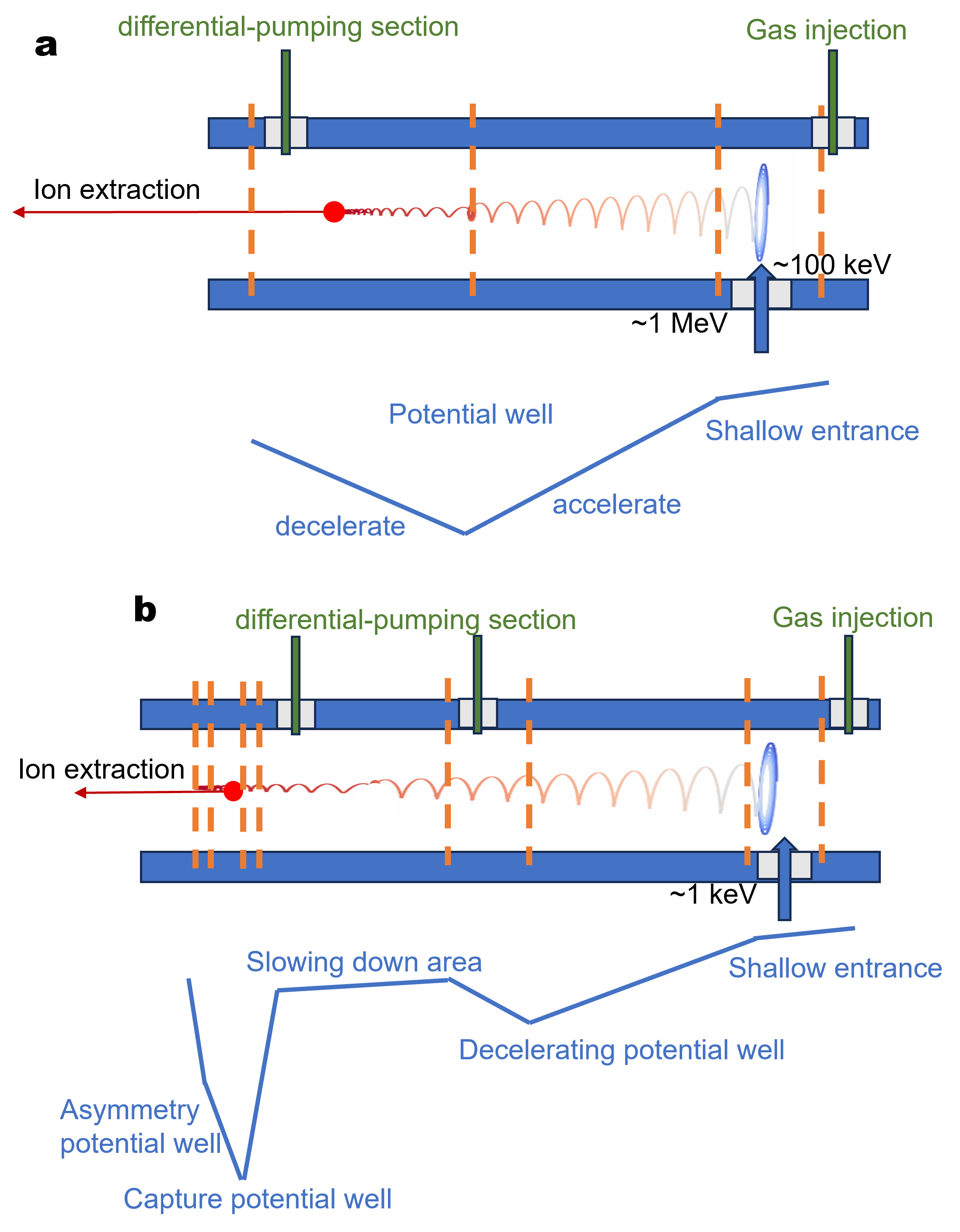}
\caption{Schematic designs of the electromagnetic gas decelerator. (a) A dedicated deceleration stage in which the muons are slowed from the initial beam energy to the desired output energy before entering
the trapping region. The axial potential is shaped to provide a broad
injection acceptance, while the gas density and potential difference
control the deceleration timescale and residual kinetic energy,
respectively. (b) A deceleration stage directly coupled to the
trapping potential. The muons are injected axially into the trapping
structure, and the transverse deceleration is completed only after
they reach approximately half of the trapping-potential depth,
reducing premature capture and losses. A shallow matching region
provides additional energy loss for muons with excessive kinetic
energy.}
\label{fig:decelerator}
\end{figure}

We therefore consider a gas-filled electromagnetic deceleration scheme based on parallel electric and magnetic fields, with the muon beam injected approximately perpendicular to both field directions. Two configurations are considered, as illustrated in Fig.~\ref{fig:decelerator}. The first is a dedicated deceleration stage, while the second is directly coupled to the trapping potential.

In the first configuration, the axial electrostatic potential is
shaped as a broad potential well, with a relatively shallow gradient
at the entrance and a steeper gradient toward the exit. The shallow
entrance region increases the acceptance for muons with a large
initial angular spread, while the gas provides continuous energy loss
during the transit. The gas density primarily determines the
deceleration timescale, whereas the potential determines
the transit time and the residual longitudinal kinetic energy at the exit. The gas pressure can be further adjusted downstream by a differential-pumping section, allowing the deceleration and extraction regions to operate at different gas densities. The gas density can be made higher in the entrance region than near the exit. Most of the deceleration can then occur upstream, while the lower density near the exit suppresses capture losses of slower muons and facilitates the transition to the subsequent high-vacuum transport section.

Near the exit, a weak high-frequency electric field can be applied
to assist the transmission of the slowest muons through the
low-energy extraction region. Because this auxiliary field is applied
only over a short distance and at low amplitude, it provides an
additional transmission mechanism without determining the overall
deceleration dynamics.

The second configuration incorporates the trapping potential directly into
the deceleration stage. To increase the injection acceptance, the muons
enter the trapping structure along its axial direction rather than
through the transverse opening. A practical difficulty is that the
effective transmission window for muons to cross the potential barrier
and enter the desired trapping region may be relatively narrow. Two
complementary measures are therefore introduced. First, a weak
high-frequency electric field can assist the transmission of the
slowest muons through the low-energy barrier region. Second, an
intermediate buffer region is placed upstream of the trapping region.
The buffer provides additional energy loss for muons that retain
excessive kinetic energy and thereby increases the effective
longitudinal acceptance of the interface. It can simultaneously serve
as a differential-pumping stage, with gas supplied on the upstream
side and pumped toward the downstream side, allowing the gas density
to remain relatively high for efficient deceleration while suppressing
gas penetration into the low-density trapping region. The potential
gradient in this region is kept shallow so that it mainly serves as an
energy- and pressure-matching region rather than introducing a strong
additional electric-field structure.

The preferred capture region is located below approximately half of the
trapping-potential depth. Muons reaching this region with sufficiently
low total kinetic energy can spend an appreciable fraction of their
axial motion near the turning points, where atomic capture is favored.
A muonic ion formed at such a position then experiences at least half
of the potential-well depth during its subsequent acceleration. This
reduces the effective charge-exchange cross section and improves the
extraction probability~\cite{LeLin2005}.

To suppress premature capture before reaching this region, the radial
motion is deliberately controlled during the upstream part of the
trapping potential. The radial deceleration is arranged to be largely
completed by the time the muons reach approximately half of the
potential depth. During the upstream half, the transverse velocity
therefore remains sufficiently high to suppress capture in the
intermediate region, where the axial velocity may already be low but
the local electric-field gradient is not yet optimal for efficient
extraction.

An important motivation for separating the deceleration into two stages
is to reduce the spread of the radial motion entering the trapping
region. The first stage removes most of the initial transverse kinetic
energy, while the second stage provides the final radial damping under
the conditions of the trapping potential. By arranging the initial
radial amplitudes entering the second stage to be as similar as
practical, the subsequent radial evolution and the location at which
the muons enter the low-velocity capture region can be made more
uniform. This reduces the fraction of muons that undergo premature
capture in unfavorable regions while retaining a broad longitudinal
acceptance.

The finite apertures required for beam injection and ion extraction also modify the idealized potential-well picture. Near the center of an aperture, the three-dimensional electrostatic potential develops a local saddle structure: the potential is higher on the aperture axis and varies more gradually in the axial direction than in the idealized one-dimensional well. Consequently, the actual minimum of the negative-muon potential is located near the electrode surface
rather than at the center of the trapping volume.

In the deceleration region, where the electrostatic field is relatively weak and the electrode spacing is large, the apertures introduce only a modest perturbation to the approximately uniform electric field. In contrast, the trapping potential has a much stronger voltage gradient over a smaller electrode spacing, so that an aperture of comparable size can produce a substantial three-dimensional field structure and act as a strong electrostatic lens rather than a small perturbation.

Because the aperture diameter is expected to be comparable to the electrode spacing, its size must balance the transverse acceptance of the incoming muon beam against the achievable electric-field strength, breakdown margin, and differential-pumping requirements. In particular, reducing the aperture substantially below the centimeter scale could impose an acceptance loss, whereas increasing it indefinitely would weaken both the electrostatic confinement and the differential-pumping capability. The optimum aperture size will therefore be determined by the three-dimensional field geometry and beam acceptance in the final design.

This three-dimensional potential structure has an important consequence for the trapping dynamics. Negative muons, as well as electrons, are therefore not expected to accumulate indefinitely at the nominal bottom of the potential well. Instead, after their transverse kinetic energy has been dissipated, they gradually approach the electrode surface and
are eventually lost. Efficient conversion thus requires the capture process to occur within the finite interval between transverse thermalization and loss to the electrode. In practice, the favorable capture region is therefore close to, but not in contact with, the electrode surface. Because the potential well is shallower on the aperture axis and deeper toward its edges, this region also benefits from a relatively large local electrostatic potential gradient, allowing a newly formed positively charged muonic ion to be accelerated efficiently after charge reversal.

The bottom of the potential well is intentionally made asymmetric to enhance the capture probability. Since capture occurs predominantly during the low-velocity intervals near the turning points of the axial motion, making the potential shallower on one side causes the $\mu^-$ particles to decelerate more gradually when approaching the corresponding turning point. They consequently spend a longer time in the low-energy region where atomic capture is favored. The steeper potential on the opposite side allows the particles to reverse their motion more rapidly. This asymmetry therefore increases the fraction of the oscillation time spent in the favorable low-energy region. The achievable degree of asymmetry is, however, constrained by the practically accessible electric-field range and must be balanced against the overall transport time.

For a typical muon beam with an energy of $100\,\mathrm{MeV}$--$1\,\mathrm{GeV}$,
the beam can first be moderated to approximately $1\,\mathrm{MeV}$
using two wedge degraders, followed by a thin Be window to reduce the
energy to about $100\,\mathrm{keV}$. The resulting low-energy muons are
then injected into the electromagnetic gas decelerator. In the first
deceleration stage, the muons are slowed to approximately
$1\,\mathrm{keV}$ and injected transversely into the second stage.
After passing the potential barrier, they become confined in the
capture potential well. Muon capture and the subsequent electronic
rearrangement produce positively charged muonic ions, which are then
accelerated out of the well by the reversed electrostatic force.

\begin{figure}[!htbp]
    \centering
    \includegraphics[width=\columnwidth]{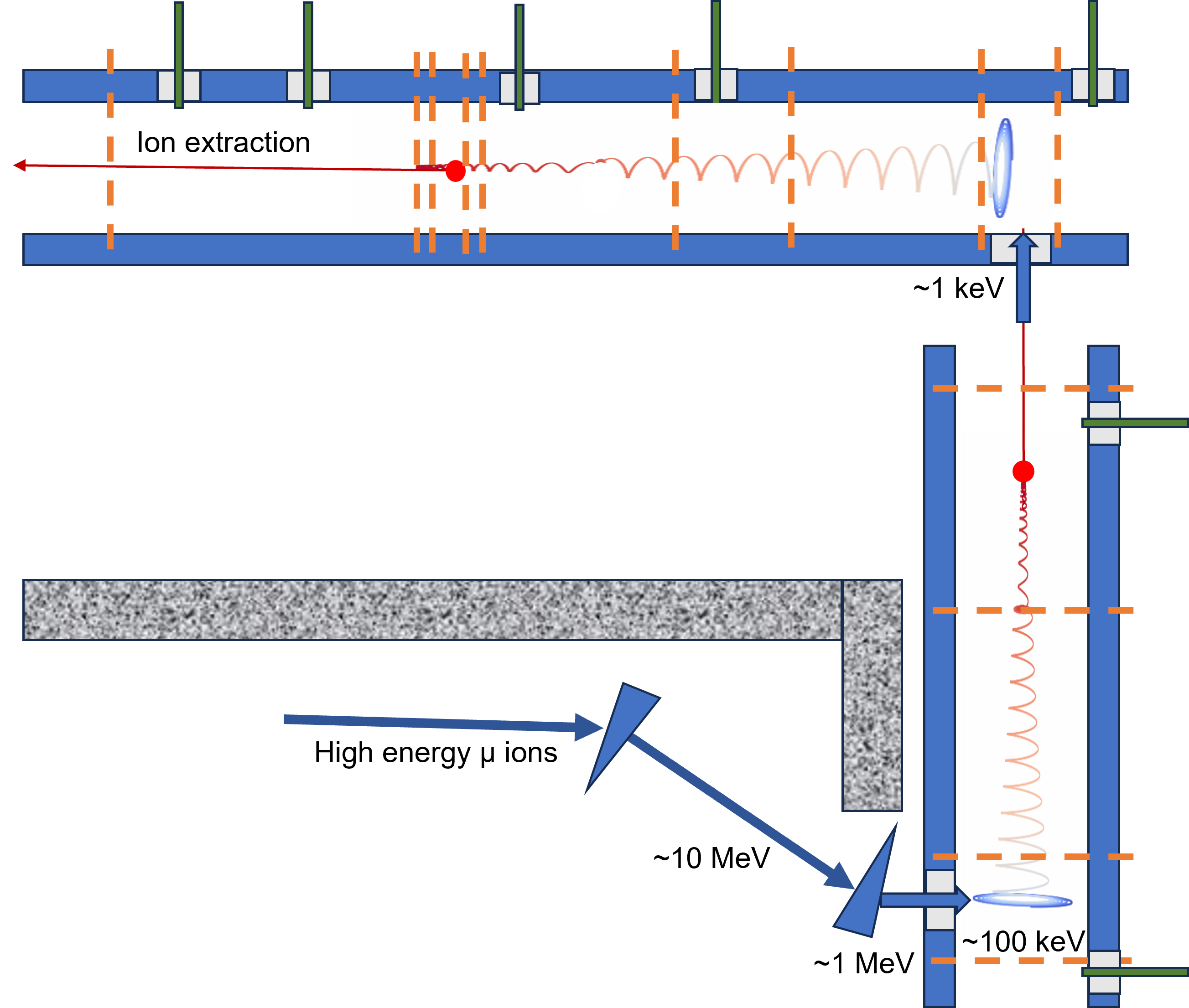}
    \caption{Schematic overview of the proposed muonic-ion extraction system. High-energy negative muons are first degraded by two wedge moderators and then pass through a thin Be window into a gas-filled electromagnetic decelerator. The decelerator slows the muons to the keV range and injects them into an electrostatic capture well, where muon capture and subsequent charge reversal produce positively charged muonic ions. These ions are accelerated out of the well by the reversed electrostatic field, pass through a differential-pumping section that reduces the gas pressure, and are finally transported into high vacuum for further acceleration by a conventional linac.}
    \label{fig:overview}
\end{figure}

After extraction, the ions enter a field-free section where
differential pumping can reduce the gas pressure before the beam is
transported into high vacuum. The extracted muonic-ion beam can then
be further accelerated by a conventional linear accelerator to
$5$--$8\,\mathrm{MeV/u}$, providing energies suitable for low-energy
nuclear-physics experiments. The complete arrangement is illustrated
schematically in Fig.~\ref{fig:overview}.

The characteristic timescales of the proposed system can be estimated from the corresponding energy-loss, capture, and acceleration processes. For the gas densities considered above, the initial deceleration from $100~\mathrm{keV}$ to the $\mathrm{keV}$ range occurs on a timescale of order $100~ns$. The subsequent buffer region is designed such that even in the limiting case of a muon entering with negligible axial kinetic energy, its transit time remains below approximately $300,\mathrm{ns}$ through the shallow electrostatic potential.

The characteristic time for atomic capture can be estimated from
\begin{equation}
\tau_{\rm cap}^{\rm th}
\simeq
\frac{1}{n\sigma_{\rm cap}^{\rm eff}v_{\rm th}},
\end{equation}
where $v_{\rm th}$ is the thermal muon velocity and
$\sigma_{\rm cap}^{\rm eff}$ represents an effective low-energy
capture cross section. At a representative density of
$n=10^{22},\mathrm{m^{-3}}$ and a temperature near liquid-nitrogen
temperature, an effective cross section of order
$10^{-18}~\mathrm{m^2}$ gives a thermal capture time of order
$10$--$100~\mathrm{ns}$. Since the muons occupy the most favorable low-velocity region for only a fraction of their axial oscillation, the effective capture time is longer by approximately a factor of $5$--$10$. We therefore estimate a characteristic capture time of order $0.1$--$1~\mu\mathrm{s}$. This estimate is intended only as an order-of-magnitude scale, since the relevant low-energy capture cross section requires dedicated atomic calculations or experimental validation.

Once charge reversal occurs, the positively charged muonic ion is accelerated by the reversed electrostatic force and rapidly leaves the capture region. For example, for a representative $\mu$Ne$^{7+}$ ion formed near half of a $50\,\mathrm{kV}$ trapping-potential depth, the remaining potential difference corresponds to an acceleration energy of approximately $175\,\mathrm{keV}$. With a mass of about $20.3\,u$, the ion reaches a velocity of approximately $1.3\times10^{6}\,\mathrm{m/s}$. For a $0.5\,\mathrm{cm}$ acceleration region, the corresponding transit time is only of order $10\,\mathrm{ns}$. The ion then traverses a $10\,\mathrm{cm}$ differential-pumping section in approximately $80\,\mathrm{ns}$, assuming that the axial electric field in this section is negligible. Thus, the total time from charge reversal to entry into the high-vacuum transport section is of order $100\,\mathrm{ns}$, and is not expected to constitute a significant contribution to the overall timescale.

Combining these contributions gives a characteristic total time of
order
\begin{equation}
t_{\rm total}
\sim
t_{\rm dec}
+t_{\rm buf}
+t_{\rm cap}
+t_{\rm acc},
\end{equation}
which is expected to remain below the muon lifetime
$\tau_\mu=2.2,\mu\mathrm{s}$ for the representative operating
conditions considered here. Thus, the finite muon lifetime does not
appear to impose a fundamental timescale barrier, although the
quantitative conversion efficiency remains sensitive to the
low-energy capture rate and the detailed phase-space evolution.

The overall conversion efficiency can be expressed schematically as
\begin{equation}
\eta_{\rm tot}
=
\eta_{\rm in}
\eta_{\rm mod}
\eta_{\rm cap}
\eta_{q}
\eta_{\rm surv}
\eta_{\rm ext},
\end{equation}
where $\eta_{\rm in}$ denotes the transmission into the
deceleration system, $\eta_{\rm mod}$ the fraction surviving
moderation and phase-space matching, $\eta_{\rm cap}$ the probability
for muon capture in the effective low-velocity region, $\eta_q$ the
population fraction of the desired muonic-ion charge state,
$\eta_{\rm surv}$ the survival probability against muon decay and
subsequent charge exchange, and $\eta_{\rm ext}$ the transmission of
the formed muonic ions through the extraction and differential-pumping
regions.

A precise prediction of $\eta_{\rm tot}$ is not yet possible because
the low-energy atomic-capture and charge-state distributions of muonic
neon require dedicated microscopic calculations or experimental input.
Nevertheless, an order-of-magnitude estimate can be constructed from
the macroscopic transport and trapping requirements. For a compact
trapping geometry matched to the low-energy muon beam, the combined
transmission through injection, moderation, and phase-space matching
may plausibly reach
\begin{equation}
\eta_{\rm in}\eta_{\rm mod}\sim 0.1\text{--}0.5.
\end{equation}
The capture probability within the effective low-velocity region is
expected to be of order
\begin{equation}
\eta_{\rm cap}\sim 0.1\text{--}0.5,
\end{equation}
depending primarily on the gas density, radial damping, and the
fraction of the axial oscillation spent at sufficiently low collision
energy. The lower end of this range allows for incomplete radial
thermalization and losses before the muons reach the preferred capture
region.

The charge-state factor $\eta_q$ is expected to represent the largest microscopic uncertainty. If the electronic cascade following muon capture efficiently removes the remaining electrons, at least one moderately highly charged state may acquire a substantial population. As a conservative reference value, we assume that a charge state with $q\gtrsim3$ can be populated with a branching fraction of order $10\%$, giving
\begin{equation}
\eta_q\sim0.1.
\end{equation}

The survival probability against muon decay and charge-changing collisions can be estimated from the total timescale discussed above. For a total processing time of approximately $0.5$--$1.5\,\mu\mathrm{s}$, we adopt a conservative range of
\begin{equation}
\eta_{\rm surv}\sim 0.3\text{--}0.7.
\end{equation}

Finally, $\eta_{\rm ext}$ includes the transmission of the formed
muonic ions through the extraction aperture and the differential-
pumping section. Because capture can occur near either end of the
trapping potential, while the two extraction directions need not have
identical transmission, we include a factor of approximately
$30$--$70\%$ in this term to account for the directional partition and
the associated aperture and transport losses.

Combining these factors gives a conservative overall conversion efficiency of order
\begin{equation}
\eta_{\rm tot}\sim10^{-4}\text{--}10^{-2},
\end{equation}
which indicates that the proposed scheme can provide an adequate
conversion efficiency for producing a useful muonic-ion beam.

The charge-state distribution $P(q)=P[(\mu A)^{q+}]$ can also be
characterized experimentally by analyzing the extracted muonic ions
after their sub-microsecond acceleration and transport. Such
measurements would directly constrain the charge-state population and
the charge-exchange losses relevant to the extraction process.

The microscopic processes governing muon capture and subsequent
charge-state evolution are not sufficiently constrained to determine
a unique optimal operating point a priori. The conversion efficiency
must therefore be optimized through experimentally accessible control
parameters. The principal parameters are the gas density, magnetic
field, and electrostatic-potential profile, which control, respectively,
the energy-loss rate, radial motion, and axial transport and extraction
dynamics.

The gas density in the trapping region is constrained by the extraction
process and should remain sufficiently low to suppress charge-changing
losses after muon capture. In the deceleration region, however, the gas
density can be varied independently through differential pumping and
provides a direct control of the transverse energy-loss timescale.
Together with the magnetic field, it can therefore be adjusted so that
the transverse kinetic energy is dissipated predominantly near the
desired position within the trapping potential.

The magnetic field confines the transverse motion and controls the
radial approach toward the aperture, while the electrostatic-potential
profile determines the axial transit time, the effective capture
region, and the subsequent acceleration of the formed muonic ions. These parameters
can therefore be varied to optimize the balance between capture
probability, charge-state survival, and extraction efficiency.

Neon is adopted here as a representative stopping gas: as a monatomic gas it avoids the additional collision and fragmentation channels associated with molecular targets, and its nuclear-capture probability is sufficiently small that muon loss during the relevant moderation and trapping times remains limited~\cite{Kirch1999,Hauser1998}. The choice is not restricted to neon, however; heavier noble gases or even molecular gases, including
volatile hydrides of elements of interest, could in principle be
used, provided their muon-capture, charge-exchange, and electron-impact
processes remain compatible with the required timescales and
electric-field conditions. This flexibility allows the scheme to be
extended to elements that are not gaseous in elemental form.

The potential impact of the proposed scheme can be illustrated by considering the intensity of currently available high-energy muon beams~\cite{Xu2025,Shimomura2024}. High-energy negative-muon beams with intensities of $10^{5}$--$10^{7} pps$ are within the range of existing and near-future accelerator-based sources, while substantially higher muon intensities are being pursued at next-generation facilities. For an overall conversion efficiency of order $10^{-4}$--$10^{-2}$, a $10^{6} pps$ input $\mu^-$ beam would correspond to an extracted muonic-ion beam of order $10^{2}$--$10^{4} pps$. With future high-intensity muon sources, the extracted intensity could consequently reach much higher values even at modest conversion efficiencies.

These estimates indicate that the proposed mechanism could
convert an experimentally relevant fraction of an intense negative-muon beam into a transportable beam of positively charged muonic ions. The availability of such a beam would open a new route to experiments on muonic atoms and muon-induced nuclear processes, while the proposed apparatus itself could provide an experimental environment for determining the microscopic quantities that currently limit the quantitative prediction of the conversion efficiency.

\begin{acknowledgments}
The author thanks ChatGPT, DeepSeek, and Kimi for extensive discussions during the development of the physical concept and experimental design presented in this work.
\end{acknowledgments}

\end{document}